\documentclass[prx,noshowpacs,print,twocolumn]{revtex4}
\usepackage{pifont}
\usepackage{amssymb}
\usepackage{amsmath}
\usepackage{graphicx}
\usepackage{dcolumn}
\usepackage{bm}
\usepackage{txfonts}
\usepackage{appendix}
\usepackage[dvips]{color}
\usepackage{bm}
\usepackage{multirow}
\begin{document}
\title{Edge and sublayer degrees of freedom for phosphorene nanoribbons with twofold-degenerate edge bands via electric field}
\author{Yi Ren}
\author{Xiaoying Zhou}
\author{Guanghui Zhou}
\email{ghzhou@hunnu.edu.cn}

\affiliation{Department of Physics, Key Laboratory for
Low-Dimensional Quantum Structures and Quantum Control (Ministry of
Education), and Synergetic Innovation Center for Quantum Effects and Applications,
Hunan Normal University, Changsha 410081, China}

\begin{abstract}

For the pristine phosphorene nanoribbons (PNRs) with edge states,
there exist two categories of edge bands near the Fermi energy ($E_F$),
i.e., the shuttle-shaped twofold-degenerate and the near-flat
simple degenerate edge bands. However, the usual experimental measurement may
not distinguish the difference between the two categories of edge bands. Here we
study the varying rule for the edge bands of PNRs under an external
electrostatic field. By using the KWANT code based on the tight-binding
approach, we find that the twofold-degenerate edge bands can be divided
into two separated shuttles until the degeneracy is completely removed and a gap near
$E_F$ is opened under a sufficiently strong in-plane electric field.
Importantly, each shuttle from the ribbon upper or lower edge
outmost atoms is identified according to the local density of states. However,
under a small off-plane field the shuttle-shaped bands are easily induced
into two near-flat bands contributed from the edge atoms of the top and
bottom sublayers, respectively. The evidence provides the edge and
sublayer degrees of freedom (DOF) for the PNRs with shuttle-shaped
edge bands, of which is obviously different from another category
PNRs intrinsically with near-flat edge bands. This is because that the former
category of ribbons solely have four zigzag-like atomic configurations
at the edges in each unit cell, which also results in that the property is robust
against the point defect in the ribbon center area.
As an application, furthermore, based on this issue we propose a homogenous junction of a
shuttle-edge-band PNR attached by two electric gates. Interestingly, the transport
property of the junction with field manipulation well reflects the
characteristics of the two DOFs. These findings may provide a further
understanding on PNRs and initiate new developments in PNR-based
electronics.
\end{abstract}

\maketitle

\section{Introduction}
The internal degree of freedom (DOF) of electrons in nanostructures
is an important issue of modern condensed matter physics. In addition
to the charge and spin DOFs, other ones have also been discussed. For
example, in multilayer graphene \cite{Novoselov} and transition-metal
dichalcogenides (TMDs) \cite{Mak}, both the layer and valley DOFs are
presented \cite{Jose, Xu}. In addition, new DOF may appear upon
tailoring a two-dimensional (2D) material into a nanoribbon. The edge
and layer, for instance, have been regarded as tunable equivalents of
the spin-one-half DOF in bilayer phosphorene nanoribbons (PNRs) with
zigzag-edge \cite{Soleimanikahnoj}. However, here we treat the edge
and sublayer as two DOFs in a recently revealed monolayer slope-edged
PNR (sPNR) with twofold-degenerate edge bands \cite{Ren}.

Phosphorene is a few- or mono-layer black phosphorus (BP) where P
atoms arranged in the top and bottom sublayers of a puckered honeycomb
lattice \cite{Li, Ling}. Inside phosphorene, each P atom is covalently
bonded with three adjacent atoms to form a puckered honeycomb structure
due to the \emph{sp}$^3$ hybridization \cite{HanCQ, Carvalho}. This
promising new 2D material, in the sense of applications in
nano-electronics, can be exfoliated from bulk BP due to the weak
interlayer Van der Waals interaction and possesses a direct band gap
of 0.3 eV \cite{Li, HanCQ}. This direct gap increases up to $\sim$2 eV
as its thickness decreases from bulk to monolayer \cite{Li1}. In the
aspect of application, the field-effect transistor (FET) \cite{Li,
Steven, Buscema, XiaF} based on phosphorene is found to have an on/off
ratio up to $10^3$ and a high hole carrier mobility to 800 cm$^2$/Vs \cite{Sherman}.
Further, arising from the low symmetric and highly anisotropic structure,
phosphorene owns strongly anisotropic electrical, thermal and optical
properties, which may open up possibilities for conceptually new
devices \cite{LiuH, HanCQ, LiPK, Zhou, Zhou1, LiuP}.

On the other hand, nanoribbons can offer better tunability in
electronic structures because of the quantum confinement and edge
morphological influence. Tailoring a 2D phosphorene sheet along the
conventional zigzag and armchair directions have been experimentally
realized \cite{Paul, Mitchell}. Hence the zigzag PNRs (zPNRs) with
significant edge states and armchair PNRs (aPNRs) with a direct
band-gap have been extensively studied \cite{Carvalho, Tran, Ezawa,ZhangR,
Ajanta}, and their skewed or beard counterparts have also been
further reported \cite{Ezawa, Marko}. In general, the edge states
projecting to the outermost atoms of a ribbon in real space are near
the Fermi level $E_F$ \cite{Ezawa, ZhouBL}. They have been extensively
studied for graphene and MoS$_2$ nanoribbons \cite{Nakada,Bollinger}.
However, a zPNR has two near-degenerate edge bands closing to $E_F$,
which are respectively contributed by the atoms of the two
edges. And the properties of these important edge states have been
recognized \cite{ZhouBL, Sousa, SK, Ren1, Sisakht, Soleimanikahnoj}.
Moreover, we may cut a phosphorene sheet so that the zigzag (armchair)
direction intersects the puckered ridges under an chiral angle other
than 0{\textordmasculine} (90\textordmasculine), resulting in PNRs
with other possible edge geometries \cite{Ren}. These
ribbons can be classified into two types, one type with edge
states including zPNR and the other without edge states including aPNR.
In specification, from our previous definition of the chiral
vector $\bm{T}$=$m\bm{a}_1/2$+$n\bm{a}_2/2$ for the crystallographic
characterizations on planar phosphorene with chiral numbers ($m,n$),
the cases of \emph{m}+\emph{n}=even integer have defined the all
possible planar crystal directions \cite{Ren}. Hence a ribbon can be generally
denoted as ($m,n$)PNR. Furthermore, according to the edge atomic
arrangement (morphology), PNRs with edge states can be further
divided into two categories. When both \emph{m} and \emph{n} are
odd, the outermost edge atoms of a PNR are alternately located at
the two sublayers, resulting in the shuttle-shaped twofold-degenerate edge bands near
$E_F$, such as (1,3)PNR and (3,1)PNR \cite{Ren, Marko,
Ashwin, Liu}. While for both even \emph{m} and \emph{n}, the
outermost atoms are located at the same sublayer and the ribbon only
has a near-flat degenerate edge band, such as (2,4)PNR and (4,2)PNR
\cite{Ren}. And some of the typical ribbons with these two
categories of edge states have already been observed in experiments
\cite{Lee, Liang}. However, the electronic property may be
significantly different from each other between the two categories.
And there are few studies on the PNRs with shuttle-shaped
twofold-degenerate edge bands \cite{Marko}. Therefore, it is
essentially demanded to explore the microscopic origin of the
twofold-degenerate edge bands, especially the transport
property. Meanwhile, the defects in PNR samples, such as monatomic
vacancies, are inevitable in experiments \cite{Kiraly}. It is also
important to understand the defect effect on the electronic
and transport properties \cite{Lill}.

In this paper we select two sPNRs, (1,3)PNR and (2,4)PNR, as the
exemplary ribbons belonging to the categories with twofold-degenerate
and near-flat degenerate edge bands, respectively.
By using the KWANT software within the framework of tight-binding method,
we find that the two shuttle-shaped twofold-degenerate edge bands of
(1,3)PNR are separated until the degeneracy
is removed and a gap near $E_F$ is opened under a sufficiently strong
in-plane electric field. And each shuttle contributed from the outmost
atoms of the ribbon upper or lower edge are identified according to the
local density of states (LDOS). However, under a small off-plane field
the shuttle-shape bands are easily separated into two degenerated
near-flat bands contributed from the edge atoms of the top or bottom
sublayer. The edge band variation with external field for this category is completely
different from that of (2,4)PNR belonging to the previously reported zPNR
category. This is because a (1,3)PNR has four zigzag atomic
configurations on the upper and lower edges and the degenerate
bands are from the outermost atoms in the same sublayer or different
upper-lower edge. This allows the two DOFs to be distinguished
and regulated by applying electric field along different directions.
Further, based on this issue we propose a (1,3)PNR homogenous junction
attached by electric gates. Interestingly, the transport property of the
junction with field manipulation well reflects the characteristics of
the two DOFs for (1,3)PNR category. In addition, the defect effect
on the transport property is also discussed. The conclusion is that both DOFs are robust
against the defect in the center area of the ribbon, but the sublayer DOF is more
effective to resist the edge vacancy than the edge DOF. These
results may provide a further understanding on PNRs and initiate new
developments in PNR-based electronics.

\begin{figure}[t]
\centering
\includegraphics[bb=67 40 737 528,width=8.4cm]{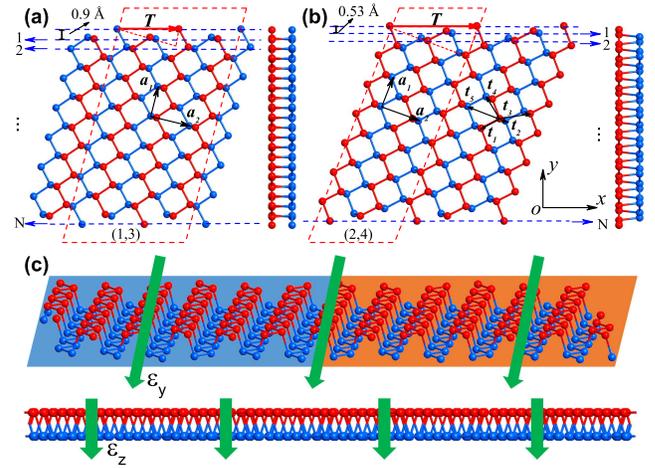}
\caption{The schematic illustrations of (a) (1,3)PNR and (b)
(2,4)PNR, where the red dashed line parallelogram in each ribbon
indicates its minimum periodical supercell, the chiral vector
\textbf{\emph{T}} is illustrated by the red solid arrow along
the edge, $\bm{a}_1$ and $\bm{a}_2$ denote the primitive
vectors, and $t_1$-$t_5$ are the five effective hoping parameters.
The distance between atomic rows for the two sPNRs is 0.9 {\AA}
and 0.53 {\AA}, respectively. And the ribbon end side views
in (a) and (b) indicate the top/bottom sublayer by the red/blue
color. (c) The sketch of the $p$-$n$ junction composed of a (1,3)PNR
where the different color shade in two sides implies the top
and bottom back-gates which adjust the Fermi level, and the
green thick arrows indicate the in-plane transverse and off-plane
vertical electric fields, respectively.}
\end{figure}

The organization of the paper is as follows. First, we classify
the categories of the PNRs with edge states, and then analyze
their edge atomic arrangements and unit cell choice In Sec. II,
we present the model description and the details on the calculations. Then in
Sec. III, we demonstrate the the edge state bands of the two exemplary
sPNRs blong to different categories and the edge band variations
under in- and off-plane electric fields, respectively. Hence according to
the coexistence of two (upper or lower edge and sublayer) DOFs
in monolayer (1,3)PNR, with which a $p$-$n$ junction is proposed and the
the important transport property is discussed. Finally, the conclusion
is briefly drawn in Sec. IV.

\section{Model Description and Method}
The schematic illustrations of the considered (1,3)PNR and (2,4)PNR
are shown in Figs. 1(a) and 1(b), respectively. The red dashed
parallelograms are the minimum supercells, where the red solid
arrows are the chiral vectors indicating the crystal
direction of the ribbon edges \cite{Ren}. The integer $N$ denotes the
number of atomic chains (blue dashed lines) across the ribbon width.
Further, for (1,3)PNR shown in Fig. 1(a), the red and blue
outmost edge atoms are arranged alternately on the two sublayers
(see the side view), as a consequence the supercell width must be
the twice of $\mid$$\bm{T}$$\mid$. In contrast, the outmost edge
atoms for (2,4)PNR are always located at one sublayer and the supercell
width is equal to $\mid$$\bm{T}$$\mid$. Moreover, as shown in Fig.
1(c), we propose a homogenous $p$-$n$ junction based on a (1,3)PNR,
where the blue and orange background imply the positive and negative
potentials provided by back-gate electric gates, respectively. And the thick
green arrows in upper and lower panels indicate the in-plane
transverse electric field along the $y$-direction and the off-plane
vertical one along the $z$-direction, respectively. The stability
of PNRs have been analyzed in our previous work \cite{Ren}.
Since the $\mid$$\bm{T}$$\mid$ of the two selected PNRs here are much less
than 20{\AA}, hence they are considered relatively stable and do not
appear obvious edge reconstruction.

We use the KWANT code based on tight-binding (TB) Hamiltonian
\cite{Groth} to calculate the electronic energy band for the
selected PNRs, and the atomistic quantum-transport simulations for
the proposed junction are based on the scattering-matrix method from
matching wavefunctions \cite{Yongjin, Zwierzycki, Tatiane}. In
comparison with the first-principles calculation \cite{Ren1}, this
approach can treat large nanostructures matching the usual
experimentally reachable sample size up to sub-100 nm scales with
better precision without large computational consumption.

The TB model Hamiltonian for phosphorene \cite{Rudenko} in the
presence of an electric field can be described as
\begin{equation}
H=\sum_{<i,j>}t_{ij}c_{i}^{\dagger}c_{j}+(el\varepsilon_{y/z}+U_{i})\sum_{i}c_{i}^{\dagger}c_{i}\;,
\end{equation}
where $c_i^{\dagger}(c_j)$ is the creation (annihilation) operator
of electron at site $i(j)$, the summation $\langle i,j\rangle$ means
over all the neighboring atomic sites with hopping integrals
$t_{ij}$, $l$ is the component of the atomic position from the
selected origin along the electric field direction,
$\varepsilon_{y/z}$ is the strength of the field along the $y/z$
direction as shown in Fig. 1(c), and U$_i$ is the impurity-induced
potential if available. It has been shown that five hopping
parameters [see Fig. 1(b)] are enough to describe the electronic
band structure of phosphorene. The values of these hopping integrals
suggested in the previous studies \cite{Rudenko} are $t_1$=-1.220
eV, $t_2$=3.665 eV, $t_3$=-0.205 eV, $t_4$=-0.105 eV and
$t_5$=-0.055 eV. Therefore, by solving the discrete
Schr\"{o}dinger equation corresponding to Hamiltonian (1) on the
proper basis for the supercell drawn by the red dashed parallelograms
in Fig. 1 and applying the Bloch theorem, the $k$-dependent
Hamiltonian for a PNR can be written as
$H(k)$=$H_{0,0}+H_{0,1}e^{ika}+H_{0,1}^{\dag}e^{-ika}$ in the form
of ($N'\times N'$) dimensional matrix. Here $N'$ is the number of
atoms in the supercell, $H_{0,0}$ is the matrix of the central cell,
$H_{0,1}$ the coupling matrix with the right-hand adjacent cell, and
$a$ is the length between two nearest-neighbor cells. Diagonalizing
this $k$-dependent Hamiltonian, we can obtain the band spectrum and
the corresponding eigen-wavefunctions. Then we can calculate the
LDOS using the following formula
\begin{equation}
\text{LDOS}(E,r)=\frac1{c\sqrt{2\pi}}\sum\limits_{n}|\Psi_n(r)|^2e^{\frac{-(E_{n}-E)^2}{2c^2}},
\end{equation}
where $c$ is broadening parameter, $\Psi_{n}(r)$ and $E_{n}$ are the
eigen-wavefunction and eigenvalue, respectively, in which $n$
denotes the energy band index and $r$ the atom position.

In addition, we mention that the description of the TB model has been
effectively used for PNRs with different edge geometries \cite{Soleimanikahnoj,Ren,LiuP,ZhangR,Marko,ZhouBL}.
We have verified the validity of TB approach by the first-principles calculation using the Atomistix Toolkit code \cite{Smidstrup}.
The result shows that the edge bands can also be near the Fermi energy and the difference between them is quantitative.
As for the influence of spin polarization, it is found that the total energy of (1,3)PNR for the spin polarization case is 8.82 meV lower
than that of non-polarized case. And for the (2,4)PNR, the energy difference is only 0.04 meV. This is well conceivable that the few-meV
energy difference will be mitigated by a reasonable finite temperature (few tens of Kelvin), i.e., the spin-polarized states will transit
into paramagnetic states in actual experiments. So that we do
not consider the effect of spin polarization on the results for the present work.

In calculating the conductance for the junction under an external
electric field, it is divided into the left electrode, the right
electrode, and the middle scattering region. The $S$ matrix can be
obtained by matching the wavefunctions at the two interfaces of
electrode/scattering-region. Once the $S$ matrix is obtained, the conductance
of the system at zero temperature can be calculated by using the
Landauer formula \cite{Datta}
\begin{equation}
G_{LR}=\frac{e^2}{h}T_{LR}=\frac{e^2}{h}\sum_{n\in L, m\in R}|S_{mn}|^{2},
\end{equation}
where L/R labels the left/right leads, $T_{LR}$ is
the transmission coefficient from lead L to lead R, and
$S_{mn}$ gives the scattering amplitude from an incoming mode $n$ to
an outgoing mode $m$, both of which are the elements of the scattering
matrix. It is clear that the conductance depends
on the number of available transport modes through the junction.

\section{Results and Discussions}

\subsection{edge bands under in-plane electric field}

\begin{figure*}
\centering
\includegraphics[bb=10 10 1553 642, width=17.5cm]{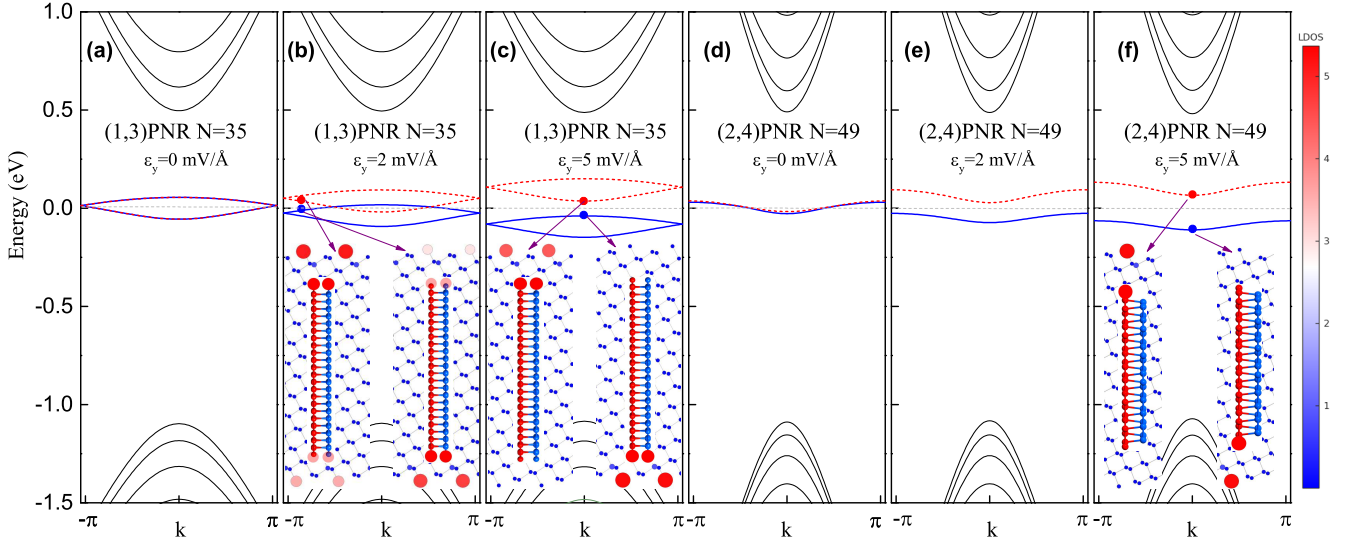}
\caption{The band structure for (a-c) (1,3)PNR with $N$=35 and (d-f)
(2,4)PNR with $N$=49 under an in-plane electric field with different
strengths, (a and d) $\varepsilon_y$=0, (b and e) 2 mV/{\AA} and
(c and f) 5 mV/\AA, respectively. The Fermi energy $E_F$ is set to zero, and
the left and right inserts in (b), (c) and (f) indicate the LDOS
distributions corresponding to energies near $E_F$ marked by
the red and blue points, respectively, where the side view for each
supercell LDOS is embodied. The right-side attached color
bar indicates the electron density from the lowest (blue) to the highest
(red) value.}
\end{figure*}

The calculated energy spectra for the selected exemplary (1,3)PNR
($N$=35) and (2,4)PNR ($N$=49) with nearly the same width $\sim$3 nm
under an in-plane electric field with different strength are
shown in Figs. 2(a-c) and 2(d-f), respectively, where the Fermi level
$E_F$ is set to zero. The left and right inserts in (b, c and f)
indicate the LDOS corresponding to energies marked respectively
by the red and blue points contributed by the ribbon upper
or lower edge outermost atoms, where the side view for each
supercell is embodied. First, as shown in Fig. 2(a), the band structure
of the pristine (1,3)PNR has a shuttle-shaped edge band near $E_F$,
which implies the ribbon is metallic. In fact, the shuttle-shaped edge bands are
twofold-degenerate differentiating by the red dashed and blue solid
lines. However, the degeneracy can be eliminated by applying an
in-plane electric field along the $y$-direction. As a small field with
strength $\varepsilon_y$=2 mV/{\AA} is applied
shown in Fig. 2(b), the twofold-degenerated edge bands obviously
separate into two partially overlapped shuttles. Interestingly, as
shown in Fig. 2(c), the two shuttles are completely separated as the
filed strength is further increased up to 5 mV/{\AA}, which means the
degeneracy is completely removed. In contrast, for the (2,4)PNR
as the example of the another category, it has two near-flat
degenerate edge state bands in the original band structure which
are drawn by the red dashed and blue solid lines shown in Fig. 2(d).
They also pass through $E_F$ exhibiting metallicity, which
is very similar to that of a conventional zPNR except for a little
difference in the curvature of the edge bands \cite{Carvalho, Ezawa,
Guo}. However, much different from the twofold-degenerated bands,
the degeneracy for this category of ribbon can be easily removed by
a small electric field as shown in Figs. 2(e) and 2(f). The
characteristic of edge band under an electric field for (2,4)PNR is
very similar to the extensively studied zPNR \cite{Ezawa, ZhouBL,
Zhang}, and hence we attribute them to the same category.

An in-plane electric field removing the degeneracy of the edge bands
can be understood from Hamiltonian (1). The additional diagonal
terms increase (accumulate) linearly with the $y$-coordinate. From
the analysis of the calculated results, it can be seen that the
degenerate edge state bands for both categories of sPNRs would be
broken by the application of an in-plane electric field. As the field
strength increases, a band gap is opened and the transition from metal to
semiconductor is occurred. The variation of (2,4)PNR edge
bands with electric field is basically the same as
zPNR \cite{Ezawa,ZhouBL, Zhang}. This is because that they belong to
the same category of sPNRs, where the outermost atoms of the ribbon
are all at the same sublayer. However, for the (1,3)PNR with
twofold-degenerate edge bands, the $E_F$ is embedded in a
mirrored shuttle-shape bands. In order to further know which atoms in the ribbon contribute the
edge state bands, we have calculated the LDOS at the energies
near $E_F$ marked by the red and blue points, respectively.
When the two shuttle-shape bands are not completely separated
shown in Fig. 2(b) for $\varepsilon_y$=2 mV/{\AA}, from the inserts
we see the difference in LDOS between energies on the two shuttles.
The red/blue dashed/solid shuttle bands are mainly contributed the outermost
atoms at the upper/lower edge.

In addition, the left and right
inserts in Figs. 2(c) and 2(f) depict the real-space electronic distributions
at the energies of conduction band minimum (CBM) and valence band maximum (VBM)
marked by the red and blue points, respectively. The right-side attached color
bar indicates the density from the lowest (blue) to highest (red) value.
From the LDOS inserts we can identify whether an edge band is completely
contributed by the outmost atoms of the upper or lower edge of the ribbon,
which provides a (upper-lower) edge DOF. However, we cannot distinguish them from
which sublayers since their responses to an in-plane field are the same.
Therefore, we need to apply an off-plane vertical electric field along
the $z$-direction respectively to the two categories of sPNRs.

\subsection{edge bands under off-plane electric field}

\begin{figure*}
\centering
\includegraphics[bb=10 10 1873 633, width=17.5cm]{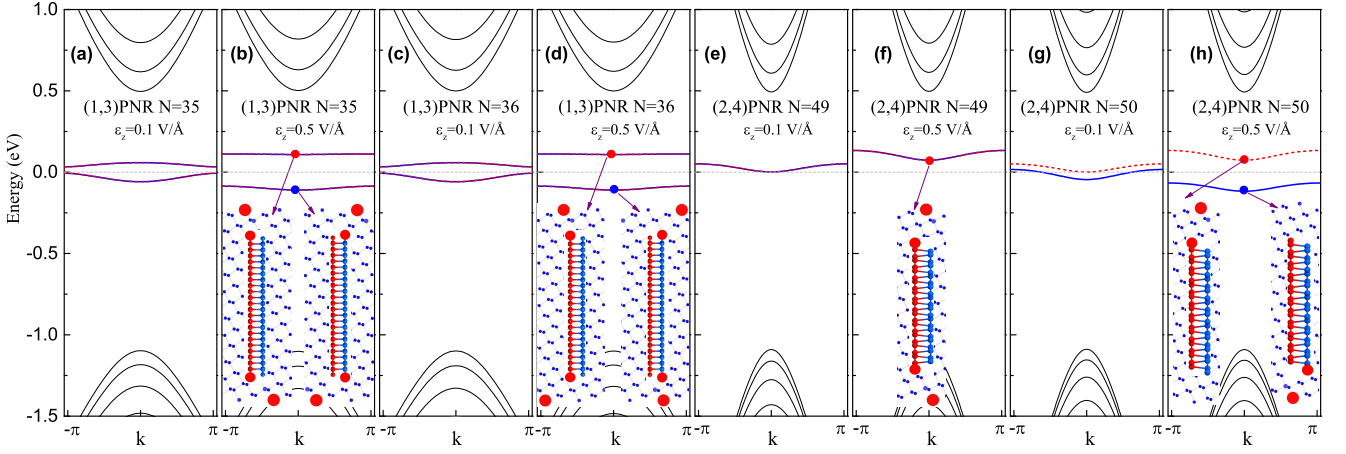}
\caption{The band structures for (a-d) (1,3)PNR with odd ($N$=35) and even
($N$=36) parities under an off-plane vertical electric field with
different strengths, respectively, where (a and c) $\varepsilon_z$=0.1 V/{\AA}
and (b and d) 0.5 V/\AA. The band structures for
odd-even ($N$=49-50) (2,4)PNR under the same field are
correspondingly shown in (e-h). The inserts in (b, d, f and h)
depict the LDOS distributions corresponding to the edge states at
the certain energies pointed by the red and blue points, where the side view for each
supercell LDOS is embodied. The color bar indicating the density is neglected here since it
is the same as that in Fig. 2.}
\end{figure*}

The difference in the layer atomic arrangement on the edge of a
sPNR with different parity rows can be reflected by the response to
an off-plane electric field, and then affects the corresponding band
structure. This effect has been revealed and discussed for
other 2D material nanoribbons, such as graphene, silicene and
phosphorene ones \cite{ZhouBL, LiZ, Kang}. Here, in Fig. 3 we present
the result of the edge band response to a vertical electric
field with strengths $\varepsilon_z$=0.1 and 0.4 V/{\AA} for
the two exemplary ribbons, respectively. For the consideration
of the even-odd parity \cite{LiuP,ZhouBL}, we consider the two
width cases of $N$=35, 36 and $N$=49, 50 for (1,3)PNR and
(2,4)PNR, respectively. The LDOS left/right insert corresponding to
the energies marked by the red/blue point. From Fig. 3(a) for
(1,3)PNR with $N$=35, the twofold-degenerate edge bands in
shuttle-shape are almost unchanged under small $\varepsilon_z$=0.1 V/{\AA}.
However, when $\varepsilon_z$ is increased up to 0.4 V/{\AA}, interestingly,
the two shuttle-shape edge bands are separated into two nearly flat
bands as shown in Fig. 3(b). This results implying a transition
from metal to semiconductor phase. But the two-fold degenerate is
still not removed. Further, from Figs. 3(a-d) it seems that the
number of atom rows (parity) does not affect the band structure
of (1,3)PNR. In addition, the side view in (b) shows that the left/right inserted LDOS come
from the edge outermost atoms of the top/bottom sublayer. We can identify
a band from the top or bottom sublayer, which provides a sublayer DOF.
This is because that the edge outmost atoms are alternately located
at the top and bottom sublayers. In the contrast, for (2,4)PNR belongs
to another category with two near-degenerate edge bands, the response of the
edge bands to the field is sensitive on the ribbon width
(parity), which is the same as that for the conventional zPNRs \cite{ZhouBL}. As shown
in Figs. 3(e) and 3(f), in specification, with the increase of the field
strength the edge band degeneracy of the odd-numbered ($N$=49)
ribbon can not be broken but enterally move upward a little. Since the
degenerate edge bands are contributed by the outermost atoms of
both edges. However, for the even-numbered ($N$=50) ribbon
shown in Figs. 3(g) and 3(h), the outermost atoms of the two edges come
from the same sublayer. Hence as the field strength increasing,
the edge band degeneracy is removed and the transition from metal to
semiconductor occurs.

\subsection{(1,3)PNR homogenous junction}
Next, from the above results for the two exemplary sPNRs belonging to
different categories, we know that a monolayer (1,3)PNR owns two
DOFs of edge and sublayer, which is particularly similar to a bilayer
zPNR with two DOFs of edge and layer \cite{Soleimanikahnoj}. Using
this similarity, we may construct a homogenous $p$-$n$ junction using
a (1,3)PNR. The two sides of a (1,3)PNR are attached by the near top
and bottom back-gates which adjust the $E_F$, leading to the
lifting-up or -down of the edge bands for two ends. Further, the external
in-plane and off-plane electric fields applied on the junction shown in
Fig. 1(c) can be realized by using a side-gate and another back-gate
electrodes, respectively. The side-gate technique has been proven to
be experimentally feasible in graphene as a channel material for other
applications \cite{Hahnlein, Molitor}. The back-gates can arouse a
potential difference (electric field) across the whole monolayer.
This kind of setup has been realized for bulk phosphorene
transistors \cite{Tayari,Kim}. And a pseudospin field effect transistor
has also been proposed and characterized based on a bilayer zPNR-based
junction, in which a pseudospin-polarized current is generated \cite{Soleimanikahnoj}.
The similar interesting effect may also be realized in the $p$-$n$
junction based on a monolayer (1,3)PNR, which can generate an
edge- or sublayer-polarized current by properly adjusting the
gate electrodes.

\begin{figure}[t]
\centering
\includegraphics[bb=9 4 596 623, width=8.5cm]{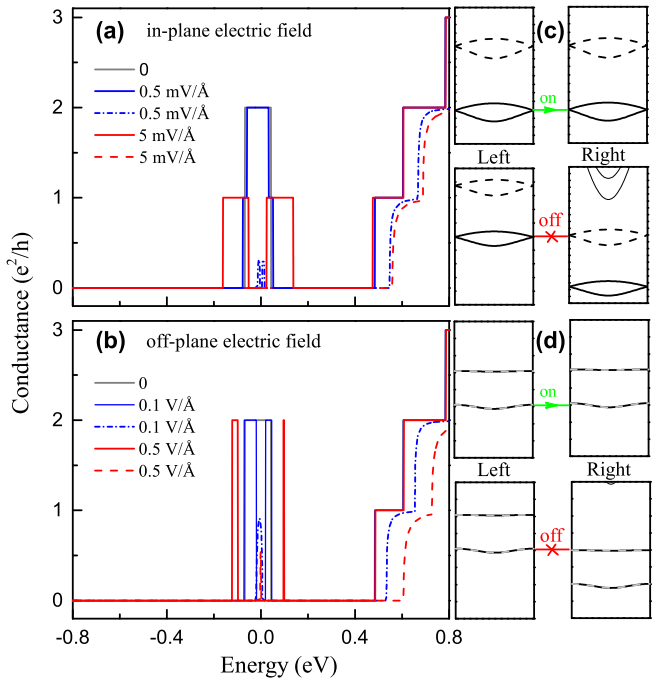}
\caption{The conductance spectrum for the (1,3)PNR-based $p$-$n$
junction with a width about 3 nm under (a) in-plane transverse and
(b) off-plane vertical electric fields, respectively. The gray
solid line implies the conductance without applied field. The blue and
red solid (dashed) lines indicate the edge bands from the
same (different) (a) upper-lower edge with field strength 0.5 and
0.5 mV/{\AA}, (b) top-bottom sublayer with field strength 0.1 and
0.5 V/{\AA} are aligned, respectively. (c) and (d) The alignment of
the edge bands of the two semi-infinitive ribbons under transverse
and vertical fields, respectively, where the green/red arrow implies
the on/off state of the junction.}
\end{figure}

In Fig. 4(a), we show the calculated conductance spectrum for the
proposed junction based on (1,3)PNR with a width $\sim$3 nm under a
transverse in-plane field with different strengths $\varepsilon_y$=0
(gray line), 0.5 (blue) and 5 mV/{\AA} (red), respectively. The solid/dashed
line indicates the alignment for the edge bands from the same/different
upper-lower edge. The separated edge bands from the two
edges of the terminals with/without alignment by
back-gates is shown in Fig. 4(c), where the upper/lower panel
corresponds to the case of the solid/dashed line in 4(a), and the
green/red arrows implies the switch on/off state. First, we find a
conductance plateau of $2e^2/h$ around $E_F$ for the
junction without external field shown by the grey solid
line, which is in accord with the band structure for a pristine (1,3)PNR
shown in Fig. 2(a). When a small field 0.5 mV/{\AA} is applied,
by adjusting the positive and negative voltages of the two back-gate
as to the edge bands from the same upper or lower edge be aligned,
the conductance still maintains a $2e^2/h$ plateau (blue
solid line) near $E_F$, whereas the conductance
is decreased to $0.3e^2/h$ from different edge as
shown by the blue dot-dashed line. This is because that the band
degeneracy is broken not enough so that the two shuttle-shaped edge bands are
not completely separated. Further, as the field strength increases
up to 5 mV/{\AA}, the degenerate shuttle-shaped edge bands is fully
removed and a significant band gap is opened around $E_F$
[see Fig. 2(c)]. As a consequence, when the
same edge bands are aligned as shown in the upper panel in 4(c), a
conductance gap correspondingly appears with two $2e^2/h$ plateaus
beside it (red solid line). Otherwise the conductance is almost
zero even though there are two shuttle-shaped edge bands from
the different upper or lower edge are aligned as shown in the lower
panel in 4(c), which shows a transport off state.

Since the speciality of the edge atom arrangement for the monolayer
(1,3)PNR, sublayer is also an important DOF in its transport property.
In Fig. 4(b), we show the conductance spectrum for the junction
under an off-plane vertical electric field with strengths,
0 (gray), 0.1 (blue) and 0.5 V/{\AA} (red), respectively. Here
the solid/dashed lines indicates the two edge bands from the
same/different sublayers are aligned, of which the diagrammatic
sketch for the edge band alignment is shown in 4(d).
First, we also find a conductance plateau of $2e^2/h$
around $E_F$ for the junction without external field
shown by the grey solid line. In the presence of applied field
$\varepsilon_z$=0.1 V/{\AA}, there is a gap opened within the
shuttle-shaped edge bands. Therefore, as the edge bands from the
same upper or lower edge are adjusted (by back-gate) to be aligned,
the conductance exhibits two steps of $2e^2/h$ besides $E_F$
(see the blue solid line), which is in accord with the band
structure shown in Fig. 3(b) or 3(d). In this case the increase of
the field strength (e.g., $\varepsilon_z$=0.5 V/{\AA}) only
result in the two narrower steps and more far away from $E_F$
(see the res solid line). On the contrary, when the edge
bands from the same upper or lower edge are not aligned,
the conductance is nearly zero except for very small peaks at
$E_F$ as shown by the blue and red dashed lines. The
reason for the results is that the edge outermost atoms of (1,3)PNR
are alternately located at the top and bottom sublayers.
Therefore, a small vertical electric field can results in wave
function superposition in different sublayers.

\begin{figure}
\centering
\includegraphics[bb=6 6 823 654, width=8.5cm]{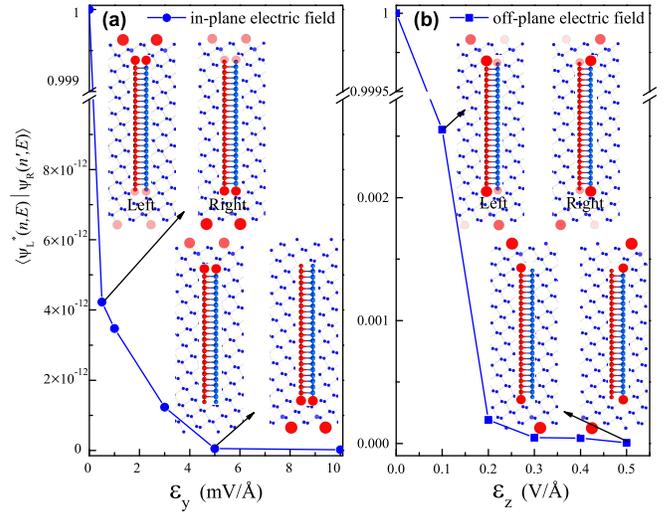}
\caption{The wavefunction overlap, $<$$\Psi_{L}^*(n,E)|\Psi_{R}(n',E)$$>$,
as a function of (a) $\varepsilon_y$ and (b) $\varepsilon_z$ for the two terminals
of the junction, where $n$ ($n'$) is the band index number for the junction left (right) side.
And the LDOS diagrams with certain energies at two different field strengths
pointed by black arrows are inserted with the embodied side view of supercell.}
\end{figure}

In order to further understand the rule of the two DOFs in transport
for the junction, in Fig. 5 we plot the dependence of wavefunction
overlap, $<$$\Psi^*_{L}(n,E)\Psi_{R}(n',E)$$>$, as a
function of the field strength (a) $\varepsilon_y$ and
(b) $\varepsilon_z$. Here $\Psi_{L/R}(n,E)$ is the wavefunction in
the left/right (L/R) side of the junction. The LDOS with certain energies
at two different field strengths are also inserted with the embodied side view of supercell.
From Fig. 5, we first note that the overlap value drops
rapidly as the field strength increasing due to
the degeneracy elimination, with which the edge bands
from the different upper-lower edge or top-bottom
sublayer are separated. Therefore, without applied field the
wavefunction overlap is equal to 1. In this case
the edge bands are completely superposed, which refers to its pristine
state shown in Fig. 2(a). Meanwhile, by adjusting the two back-gates as to
the two side edge bands from the same edge or sublayer
aligned as shown in upper panels of Fig. 4(c) or 4(d), the overlap
degree is also equal to 1 regardless of the electric field strength.

Further, as shown in Fig. 5(a), we find that the two shuttle-shaped
edge bands from the upper or lower edge are aligned by adjusting
the back-gates as to the potential difference between the junction two sides
is equal to the electric field strength [see the lower panel in Fig. 4(c)],
the wavefunction overlap for two terminals is greatly reduced. When
the two shuttle-shape edge bands are just separated a little by a small
field 0.5 mV/{\AA}, the wavefunction overlap is greatly reduced
to a small value, which corresponds to the blue dashed
peaks near $E_F$ in Fig. 4(a). This is because that the overall degeneracy of
the two shuttle bands would be almost removed even under a small electric field,
but some of the degeneracy is still remained. Therefore, at certain energies the
superposition of the wavefunctions in two sides still has a definitive value.
In this case, from the LDOS shown in left upper panel of Fig. 5(a)], we can see it
distributes at the upper and lower edge outermost atoms. Specifically, the upper edge
atoms in the left side account for the major contribution (bright red spots),
while the reddish ones at lower edge indicate a partial contribution.
On the contrast, the LDOS distributed are the upper-lower edge for
the right side is opposite to the left one. As the field strength
is increased up to 5 mV/{\AA}, the wavefunctions overlap
for two side approaches zero (as red dashed line shown in Fig. 4(a)).
This is because that the two shuttle edge bands are completely separated,
and electrons are localized only at the outmost atoms of one edge. This can also be seen
from the LDOS inserts with the embodied side view of supercell. On the other hand, since the
edge mostout atoms of (1,3)PNR are alternately arranged on the top and bottom sublayers,
the wavefunctions of the top and bottom sublayer atoms still mix
together when a large vertical electric field is applied as shown in the lower
insert of Fig. 5(b). Hence the amplitude scale of $y$-axis in Fig. 5(b) is much larger
than that in 5(a). Therefore, when $\varepsilon_z$=0.1 V/{\AA} the wavefunction
overlap for two sides is still large, which corresponds to the blue
dashed plateau near $E_F$ in Fig. 4(b). When the vertical field becomes
larger, the wavefunction overlap of the top and bottom sublayers
tends to zero due to the localization of electrons, as shown
by the lower insert on Fig. 5(b) for LDOS with 0.5 V/{\AA}.

\begin{figure}
\centering
\includegraphics[bb=10 5 517 898,width=8.0cm]{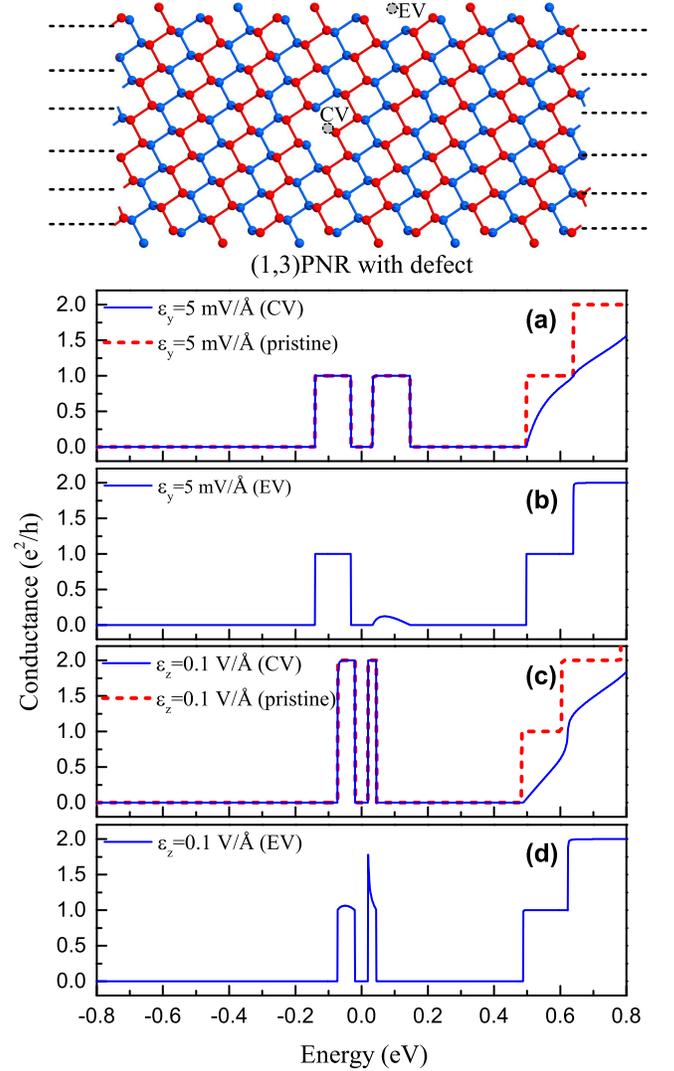}
\caption{The conductance spectrum of a (1,3)PNR with monatomic vacancy
respectively located at the (a, c) center and (b, d) edge of the ribbon, where (a, c)
with $\varepsilon_y$=5 mV/{\AA} and (c, d) $\varepsilon_z$=0.1 V/{\AA}. The red
dashed line in (a) and (c) corresponds to the pristine case without defect.}
\end{figure}

Finally, we demonstrate the influence of defect on the transport,
which is also very important for device application by using the proposed
DOFS since defects are inevitable on PNR samples in experiments \cite{Kiraly}. For the
sake of argument on the two DOFs in a (1,3)PNR, here we mainly consider
two kinds of monatomic vacancies. As shown in the top of Fig. 6 for
a defective (1,3)PNR, there are two monatomic vacancy positions indicated by
the black dotted circles, one at the edge (EV) and the other in
the center (CV) of the ribbon. Figure 6 gives the conductance spectrum,
where (a, c) and (b, d) for CV and EV cases with $\varepsilon_y$=5 mV/{\AA}
and $\varepsilon_z$=0.1 V/{\AA}, respectively. As a reference, the red
dashed lines in (a, c) indicate the conductances for a pristine (1,3)PNR without
defect but with in- and off-plane fields, respectively, which can be refer
to the corresponding energy spectrum shown in Figs. 2 and 3. As is seen from the
conductance spectra shown by the blue solid lines, a CV does not affect
the edge bands regardless the field direction, resulting in an unchanged
conductance in comparison the pristine case shown in (a, c). But the situation
for EV is much different. Comparing with the conductance spectrum shown by the red
solid line in Fig. 4(a) for the same upper-lower edge are aligned (also shown
by red dashed line in Fig. 6(a)). As shown in Fig. 6(b), we find
that the edge state on one of edges with an EV is severely weakened,
which result in that the conductance plateau on the right side of $E_F$
almost disappears. The original conductance plateau is maintained. In addition, for a
EV with $\varepsilon_z$=0.1 V/{\AA} shown by the blue solid line in Fig. 6(d),
the conductance steps are retained as the ideal case [see the blue solid
line in Fig. 4(b) and the red dashed line in Fig. 6(c)]. But the
conductance peak on the left side of $E_F$ is reduced by half,
while the right side peak is still close to $2e^2/h$. This importantly implies
that the CV has no effect on the edge state. The same result is expected
for the in-plane field case. Moreover, due to the particularity of
the two DOFs, the effect of EV on edge-state contributed conductance is
different under the regulation of electric field. As shown by the LDOS
in Fig. 5(a), a very small in-plane field can make the outer electrons
of the edge atoms localized, which provides the transport channel
corresponding to the edge with an EV almost forbidden. On the other
hand, a large off-plane field would cause a certain degree of wavefunction
overlap between the sublayer atoms on the same side [as shown in Fig. 5(b)],
so that an EV in one sublayer only destroys the edge state of the corresponding
sublayer, while the other sublayer on the same side still maintains a channel.
In general, the sPNRs with two-DOFs can reduce the influence of low-concentration
edge monatomic defects on the edge states to a certain extent.

At last but not least, besides the atomic defects the edge passivation
of nanoribbons is also an important issue because that the edges are usually
passivated in real experimental samples. When edge hydrogen(H)-passivation is considered in the TB
calculation, usually an extra potential field is applied on the edge atoms
of the ribbons. This may lead to that the original edge bands are disappeared or
modified (see, e.g., \cite{LiuP}). Here we have also tested the H- and O-passivation
cases by using a DFT calculation, respectively. The result shows that the H-passivated edges
do not exhibit edge states, while O-passivated ones remain qualitatively similar to the the case of
bare edges. And the result of the edge bands is the qualitatively the same as those by the TB approach,
which confirms that the device assumption based on (1,3)PNR is scientifically valuable \cite{Ren}.

\section{Summary and Conclusion}
In summary, we have studied the difference in electronic structures between
the two categories of sPNRs with edge states and the variation under an
external electrostatic field. Taking (1,3)PNR and (2,4)PNR as the
examples, we first identify the distinction of edge morphology and the unit
cell selection for these two ribbons. And then, by using the KWANT code
based on TB approach, we find that the shuttle-shaped twofold-degenerate edge bands of
(1,3)PNR can be became two separated shuttles until the degeneracy
is removed and a gap near $E_F$ is opened under a sufficient strong
in-plane electric field. And each shuttle contributed from the outmost
atoms of the ribbon upper or lower edge is verified according to the
LDOS. However, under a small off-plane field the shuttle-shape bands
are easily separated into two degenerated near-flat bands contributed
from the edge atoms of the top or bottom sublayer. The edge band
variation with external field for this category is completely
different from that of (2,4)PNR belonging to previously reported zPNR
category. This is because a (1,3)PNR has four zigzag atomic
configurations on the upper and lower edges and the degenerate
bands are from the outermost atoms in the same sublayer or different
upper-lower edge. This allows the two DOFs to be distinguished
and regulated by applying electric field along different directions.
Further, based on this issue we propose a (1,3)PNR homogenous junction
attached by electric gates. Interestingly, the transport property of the
junction with field manipulation well reflects the characteristics of
the two DOFs for (1,3)PNR category. In addition, the defect effect
from the vacancies in edge and bulk on the transport property is also
discussed. These results may provide
a further understanding on PNRs and initiate new developments
in PNR-based electronics.

\begin{acknowledgments}
This work was supported by the National Natural Science Foundation
of China (Grant Nos. 11774085, 11804092, and 11664019), the
Project Funded by China Postdoctoral Science Foundation (Grant Nos.
BX20180097, 2019M652777), and Hunan Provincial Natural Science
Foundation of China (Grant No. 2019JJ40187).
\end{acknowledgments}

\end{document}